\documentclass[preprint,12pt]{elsarticle}
\usepackage{graphicx}
\usepackage{amssymb}
\usepackage{amsthm}
\usepackage{epstopdf}
\usepackage{caption}
\usepackage{subcaption}

\usepackage{url}

\journal{Epidemics}

\begin{document}
\begin{frontmatter}
\title{Future of COVID-19 in Italy: A mathematical perspective}

\author{Sumit Kumar}

\address{School of Basic Sciences, Indian Institute of Technology Mandi, Mandi, Himachal Pradesh 175001, India}
\author{Sandeep Sharma}
\address{Department of Mathematics, DIT University, Dehradun, Uttrakhand, 248009}
\author{Nitu Kumari\corref{cor1}}
\ead{nitu@iitmandi.ac.in}
\cortext[cor1]{Corresponding author}
\address{School of Basic Sciences, Indian Institute of Technology Mandi, Mandi, Himachal Pradesh 175001, India}

\begin{abstract}
We have proposed a SEIR  compartmental mathematical model. The prime objective of this study is to analyse and forecast the pandemic in Italy for the upcoming months. The basic reproduction number has been calculated. Based on the current situation in Italy, in this paper, we will estimate the possible time for the end of the pandemic in the country. The impact of lock down and rapid isolation on the spread of the pandemic are also discussed. Further, we have  studied four of the most pandemic affected regions in Italy. Using the proposed model, a prediction has been made about the duration of pandemic in these regions. The variation in the basic reproduction number corresponding to the sensitive parameters of the model is also examined.
\end{abstract}
\end{frontmatter}

\section{Introduction}
Since its first appearance in Wuhan (China), the severe acute respiratory syndrome coronavirus-2 disease (COVID 19) is spreading rapidly across the globe \cite{lai2020severe,lupia20202019}. The number of patients are increasing exponentially and in some of the countries, thousands of people are losing their lives, almost every day, due to COVID 19. The seriousness of the situation can be understood by the fact that the number of infected cases have crossed the  1.2 million mark and more than 75 thousand confirmed deaths, across the globe, due to the disease as on 7th April \cite{who:2020}. Moreover, the outbreak has spread in more than 200 countries \cite{who:2020}. Owing to the grievous condition, the World Health Organization first declared the COVID 19 as a Public Health Emergency of International Concern on 30 January 2020 and later a pandemic on 11 March 2020 \cite{who1:2020}. In fact, at this point of time, the COVID 19 seems to be unstoppable. Different countries and health agencies are working together to find some robust method to prevent the ongoing wave, which now becomes a pandemic, of the disease. Moreover, the ongoing outbreak forced various countries to implement the countrywide lock down to break the chain of transmission of the corona virus. This results in a huge loss in terms of economy and resources and at the same time created a chaotic situation worldwide. Efforts have been made from different corners of the research community to understand the structure of the virus and the transmission mechanism of the disease. Different studies analyze the role of social distancing (which is the only effective method available till date) in the prevention of COVID 19. 

At the initial stage, it is believed that transmission of the disease took place through animal to human mode. But later it has been established that the direct transmission of the disease is  also possible and is the primary reason for the acute transmission to various countries \cite{li2020early,wang2020clinical,carlos2020novel}.  The possibility of hospital related transmission was also explored and it was suspected to be the possible cause in 41 \% of patients \cite{wang2020clinical}. Along with its high transmission efficiency, the high level of global travel contributed heavily to the spread of SARS-CoV-2 across the globe \cite{biscayart2020next}. 

The ongoing pandemic of COVID-19 challenged many developed nations including Germany, Spain, USA, Italy, France and several others and has caused catastrophic impacts on the lifestyle of the people of these countries. This is almost opposite to the pattern of infectious diseases observed to date. In the past, such an outbreak of pandemic generally took place in underdeveloped and developing countries. The African region which is the host of many infectious diseases has the lowest reported cases of COVID-19 till date. 

Italy recorded its first case of COVID19 on  February 20, 2020, at Lodi (Lombardy) \cite{grasselli2020critical}. In the next 24 hours, the infected cases increased to 36 \cite{grasselli2020critical}. Moreover, at the initial stage of the outbreak, Italian data followed closely the exponential growth trend observed in Hubei Province, China. Till date, Italy is one of the countries that have faced grievous consequences of COVID 19. Till 7th April 2020, Italy has 1,32,547 recorded cases and 16,523 deaths due to COVID 19. In terms of reported cases, it is the third highest while it is on the top when it comes to the number of deaths across the globe. Of the patients who died, 42.2 \% were aged 80-89 years, 32.4 \% were aged 70-79 years, 8.4 \% were aged 60-69 years, and 2.8 \% were aged 50-59 years. The male to female ratio is 80 \% to 20 \% with older median age for women (83.4 years for women vs 79.9 years for men) \cite{remuzzi2020covid}. Moreover, the estimated mean age of those who lost their lives in Italy was 81 years \cite{remuzzi2020covid}. The COVID19 outbreak completely disturbed the economic condition of Italy. Several family owned small sectors are suffering \cite{lazzerini2020covid}. The condition of Italy surprised the research fraternity because Italy stands in the top five countries in terms of medical facilities. The ongoing dismal scenario in Italy has forced the government to admit that they do not have any control over the spread of the disease as well as do not know when the ongoing web of COVID 19 will stop. 

Due to the catastrophic impacts of the COVID-19 outbreak, efforts have been made to analyze the trend of the disease and predict the future of the epidemics \cite{remuzzi2020covid,vattay2020predicting}. The work carried out in \cite{remuzzi2020covid} predicts that in the absence of timely implementation of available medical resources, the authorities will not be able to control the outbreak of the disease. The study further concludes that together with the medical facilities people’s movement and social activities should be restricted immediately in order to curtail the burden of the COVID-19. The work carried out in \cite{vattay2020predicting}, collected the day to day data and measured the possible similarity between Italy and Hubei Province (China). Further, the study also shows that the number of deaths increased almost five times as the available treatment facilities reached  the limit. The limited availability of medical staff and facilities also delays the possible end of the COVID-19 crisis from 15 April 2020 to 8 May 2020.

However, to the best of our knowledge so far no compartmental deterministic mathematical model has been developed and studied for COVID-19 spread in Italy. In the current work, we propose a compartmental mathematical model to study the case of Italy. The proposed model incorporates four different compartments namely - Susceptible, Exposed, Infected and Recovered population. The present COVID-19 has an incubation period ranges up to 14 days. Therefore, to make our model realistic, we include the exposed population along with the  infected population, which certainly results in improved prediction. We collected the data of Italy for COVID 19 available on the Worldometers\cite{worldometer} website. Further we trained the proposed mathematical model using the data avilable till $6^{th}$ April 2020. Using this improved mathematical model we study the impact of lockdown on the spread of COVID 19 in Italy. Also, we predict the possible end of the current outbreak of COVID 19 in Italy. Moreover, to make our predictions more realistic, we have trained and validated our model with COVID-19 data of some the highly affected regions of Italy.
\section{The Mathematical Model}
\begin{figure}[h]
\centering
\includegraphics[scale=0.4]{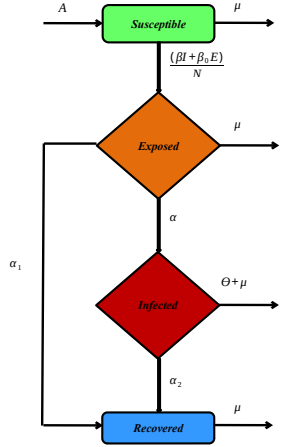}
\caption{Flow chart for the COVID-19 Italy model.}
\label{Fig_1}
\end{figure}

The proposed model describes the transmission mechanism of COVID-19. In the modelling process, we have divided the human population into four mutually exclusive compartments, namely, susceptible ($S$), exposed ($E$), infected ($I$) and recovered ($R$). The exposed class is the collection of those individuals which have just got infected and are not showing any symptoms of the disease, while the infected class individuals have clear symptoms of COVID-19. We further assume that a recovered individual will become immune to the infection. The flow chart of the model is presented in figure \ref{Fig_1}. \\
\\
Based on the above assumptions, the model is governed by the following system of equations:-

\begin{equation}
\begin{array}{l} \vspace{0.3cm}
\frac{dS}{dt}=A-\beta S \frac{I}{N}-\beta_0 S\frac{E}{N}-  \mu S  \\
\vspace{0.3cm} 
 \frac{dE}{dt}=\beta S \frac{I}{N}+\beta_0 S\frac{E}{N}-\alpha E-\alpha_1 E-\mu E \\
 \vspace{0.3cm}
 \frac{dI}{dt}=\alpha E -\theta I-\alpha_2 I -\mu I \\
 \frac{dR}{dt}=\alpha_1 E+\alpha_2 I-\mu R 
\end{array}
\label{eq_1}
\end{equation}\\

In the above model, $N$ represents the total population. We assume that all the new recruiters joined the susceptible class at a constant rate $A$. $\beta$ is the disease transmission rate from the infected individuals to  susceptible individuals. We further assume that susceptible individuals once come into the contact of infected individual will not directly join the infected class. They first join the exposed class ($E$) and after certain period of time shows visible symptoms of the disease and enters into the infected class ($I$). Exposed class individuals are assumed to be less infectious as compared to the infected class individuals. Therefore, $\beta_0$ represents the disease transmission rate for exposed individuals. Clearly, $\beta_0 \le \beta $. Here, $\alpha$ is the rate at which exposed individuals join the infected class. $\alpha_1$ is the recovery rate of exposed individuals and $\alpha_2$ is the recovery rate of infected individuals. $\theta$ is the disease induced death rate. $\mu$ is the natural death rate. 
\section{Basic Properties}
In this section, we check the mathematical feasibility of the proposed model. For this purpose, we check whether all the solutions of the proposed model will remain positive and bounded or not.

Our proposed model system involves human population. Hence for the initial state, all the compartmental values are assumed to be non-negative. We consider the following initial condition for analysis:
\begin{equation}
S(0)\ge 0, E(0)\ge 0, I(0)\ge 0, R(0)\ge 0
\label{eq_2}
\end{equation}
\subsection{Positivity of the Solution}
To show the epidemiological feasibility of the proposed model system \ref{eq_1}, it is required that all the solutions remain non-negative. Hence, in the following theorem we verify that all the solutions with non-negative initial condition will remain non-negative. The following theorem establishes the positivity of the solutions.\\\\
\textbf{Theorem 1:} The solution $\left( S(t), E(t), I(t), R(t) \right)$  of the proposed model system is non-negative for all $t \ge 0$ with non-negative initial condition.\\
\textbf{Proof :} 
From the first equation of system (\ref{eq_1}), we have\\
\begin{equation}
\begin{array}{l}
\frac{dS}{dt}\ge\frac{A}{2}-\left(\frac{\beta I}{N}+\frac{\beta_0 E}{N}+\mu\right)S 
\end{array}
\end{equation}

From this equation, we can deduce that\\ 

\begin{equation}
\begin{array}{r} \vspace{.2cm}
\frac{d}{dt}\left[S(t) \exp\left\{\int^t_0\left(\frac{\beta I(v)}{N}+\frac{\beta_0 E(v)}{N}\right).dv+\mu t\right\}\right]  \vspace{0.5cm} \ge \\ \frac{A}{2}\exp\left\{\int^t_0\left(\frac{\beta I(v)}{N}+\frac{\beta_0 E(v)}{N} \right).dv +\mu t\right\} 
\end{array}
\end{equation}
Now, integrating the equation on both sides
\begin{equation}
\begin{array}{r} \vspace{0.3cm}
S(t_1) \exp\left\{\int^{t_1}_0\left(\frac{\beta I(v)}{N}+\frac{\beta_0 E(v)}{N}\right).dv+\mu t_1\right\}-S(0) \ge \\ \int^{t_1}_0 \frac{A}{2}\exp\left\{\int^u_0\left(\frac{\beta I(v)}{N}+\frac{\beta_0 E(v)}{N}\right).dv+\mu t\right\}.du 

\end{array}
\end{equation}

On simplification we have,
\begin{equation}
\begin{array}{l} \vspace{0.3cm}
S(t_1) \ge S(0) \exp\left\{-\left(\int^{t_1}_0\left(\frac{\beta I(v)}{N}+\frac{\beta_0 E(v)}{N}\right).dv+\mu t_1 \right)\right\}+\\ \exp \left\{-\left(\int^{t_1}_0\left(\frac{\beta I(v)}{N}+\frac{\beta_0 E(v)}{N}\right).dv+\mu t_1 \right)\right\}\times \\ \\ \int^{t_1}_0 \frac{A}{2}\exp\left\{\int^u_0\left(\frac{\beta I(v)}{N}+\frac{\beta_0 E(v)}{N}\right).dv+\mu t\right\}.du \nonumber
\end{array}
\end{equation} 
This gives
\[S(t_1) \ge 0\] where $t_1 \ge 0$ is arbitrary.
Similarly we can show the non-negativity for compartments $E$, $I$ and $R$.\\ Hence, the solution $\left( S(t), E(t), I(t), R(t) \right)$ will remain positive for non-negative initial condition.
\subsection{Boundedness of the Solution}
In order to accurately predict the epidemic, the solutions of the mathematical model should be bounded. The following theorem guarantees the boundedness of the solutionsof the COVID-19 model designed for Italy.\\\\ 
\textbf{Theorem 2:} All solutions of the proposed model are bounded.\\
\textbf{Proof :} We need to show that $ \left( S(t), E(t), I(t), R(t) \right)$ is bounded for each value of $ t \ge 0$. From our model system \ref{eq_1} we obtain:
\[{\left(S+E+I+R\right)}'= A-\mu \left(S+E+I+R\right)-\theta I \le A-\mu \left(S+E+I+R\right)\]
which gives us
\[ \lim_{t \to \infty}Sup\left(S+E+I+R\right) \le \frac{A}{\mu} \]
Also, from the first equation of model system \ref{eq_1}, we have 
\[ \frac{dS}{dt}\le A-\mu S\]
From here, we can conclude that
\[ S(t) \le \frac{A}{\mu}\]
Similarly, we can show the boundedness of every compartment of the model. Therefore feasible region for our model system will be 
\[\left\{ \left( S(t), E(t), I(t), R(t) \right)| S
+E+I+R \le \frac{A}{\mu}, 0\le S(t),E(t), I(t), R(t) \le \frac{A}{\mu} \right\}\]

\section{COVID-19 in Italy}
As of now, Italy is one of those countries which are highly effected by COVID-19. According to the data collected from \cite{worldometer}, the total number of COVID-19 cases has crossed 139422 as on $9^{th}$ April 2020. There has been more then 17669 deaths in the country due to this epidemic.  Our proposed model aims to predict the future scenario of the COVID-19 epidemic in Italy by analyzing its present state in the country. In this section, we perform rigorous numerical simulations to get an insight of the epidemic in Italy. The parameter values used for the simulation are provided in Table \ref{Table: 1}. We have used all the real time parameter values avalible on the WHO website \cite{who:2020} and  \cite{worldometer}. However those parameter whose values are not available in the real time, has been assumed for the study. All the simulations are performed in MATLAB 2013a.

\begin{table}[h]
\centering
\begin{tabular}{|p{1 in}|p{1 in}|p{1 in}|p{1 in}|} \hline
Parameter & Value & Range & Source\\ \hline
$A$ & 10000 & Constant & Assumed \\ \hline
$\beta$ & 0.82 & 0.1-0.99& Assumed \\ \hline
$\beta_0$ & 0.5 & 0.1-0.99& Assumed \\ \hline
$\alpha$ & 0.7 & 0.1-0.99& Assumed \\ \hline
$\alpha_1$ & 0.2 & 0.1-0.8& Assumed \\ \hline
$\alpha_2$ & 0.6 &0.1-0.8& Assumed \\ \hline
$\theta$ & 0.17 & 0.1-0.99 & \cite{worldometer}\\ \hline
$\mu$ & 0.00002904 & Constant & \cite{Knoema}\\ \hline
$ N$ & 60500000 & Constant & \cite{worldometer}\\ \hline
\end{tabular}
\caption{Parameter values used for the model}
\label{Table: 1}
\end{table}

We have considered the following initial condition to perform our analysis. This is taken on the basis of the officially reported data by \cite{worldometer} as on $1^{st}$ March 2020.
\[\left( S(0), E(0), I(0), R(0)\right)=\left(50000000, 18000, 1577, 83 \right)\]
According to the situation report-11  available on the official website of WHO\cite{situation}, the first two COVID-19 positive cases in Italy were reported on $31^{st}$ January 2020. Both the infected individuals had travel history to the city of Wuhan, china. This was the initial phase of the epidemic in the country. Due to the lack of government concern on this matter, the epidemic started to grow slowly but significantly. By the end of February the number of cases reached to 1000 \cite{worldometer}. We trained our model defined for Italy for the month of March to check its accuracy with the official data available \cite{worldometer}. It was observed that our model predicted the epidemic closely. Figure \ref{fig_2} shows the measured data by \cite{worldometer} and the model simulation for the possible number of COVID-19 cases in Italy. \\
The goal of this study is to estimate the scenario of the epidemic for the upcoming days in Italy. For this purpose, we perform numerical simulation and fit the proposed model to the real data available till $31^{st}$ March 2020.

\begin{center}
\begin{figure} [h] 
\includegraphics [width=.8\textwidth]{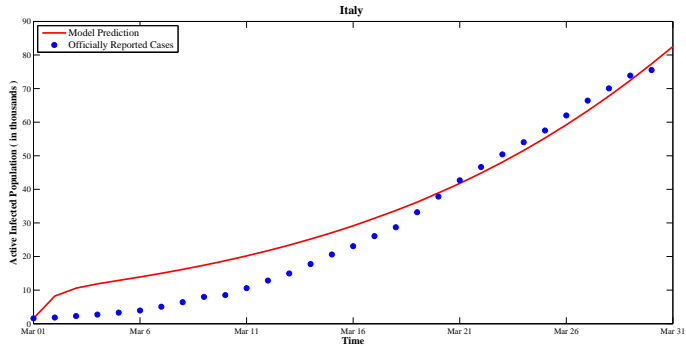}
\caption{The model has been fitted for the COVID-19 outbreak in Italy for the month of March-2020. The blue dots are the officially reported data and the solid red line represents the infections predicted from proposed model.}
\end{figure}
\label{fig_2}
\end{center}

\begin{center}
\begin{figure} [h]
\includegraphics[width=.8\textwidth]{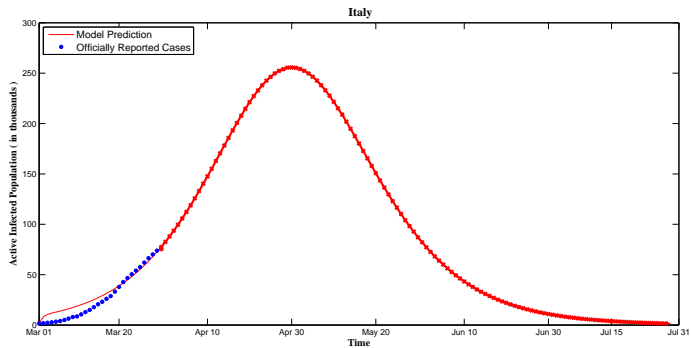}
\caption{Model prediction of COVID-19 in Italy and its possibility to die out in July, 2020.}
\label{fig_3}
\end{figure}
\end{center}
 It can be seen from the simulated results (see fig.\ref{fig_3}) that, the active cases will still increase till the end of April. The pandemic cases may start to decrease by the first week of May. The epidemic may hit its peak in Italy between $28^{th}$ April to $3^{rd}$ May as shown in figure \ref{fig_3}. Also, according to our proposed model and if the conditions remains the same, the epidemic in the country may not last till the end of July 2020.

 \section{Game Changers}
In this section, we will discuss the factors which can significantly control the spread of pandemic. The two major factors discussed here are (a) Early Lock down and (b) Rapid Isolation. 
\subsubsection{Impact of Early Lock down}
On $9^{th}$ March 2020, the government of Italy imposed a nationwide quarantine restricting the movement of the population except for necessity, work, and health circumstances, in response to the growing pandemic of COVID-19 in the country. In this section, we will investigate the effect of early lock down on the spread of pandemic in Italy.
\begin{figure}[h]
\begin{subfigure}{.5\textwidth}
\centering
\includegraphics[width=\linewidth]{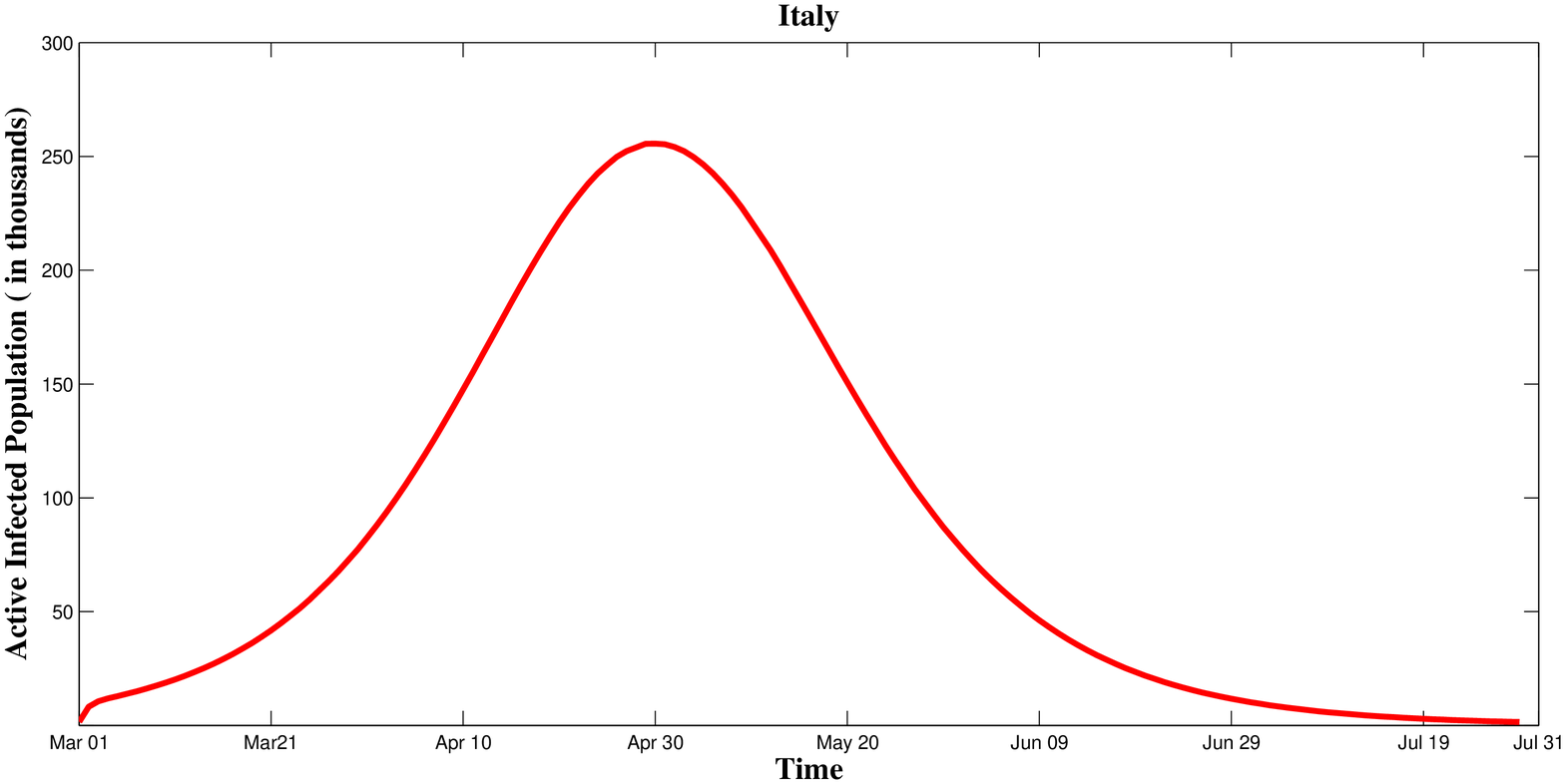}
\caption{$S(0) = 5$ Crores}
\label{fig:sub_4a}
\end{subfigure}
\begin{subfigure}{.5\textwidth}
\centering
\includegraphics[width=\linewidth]{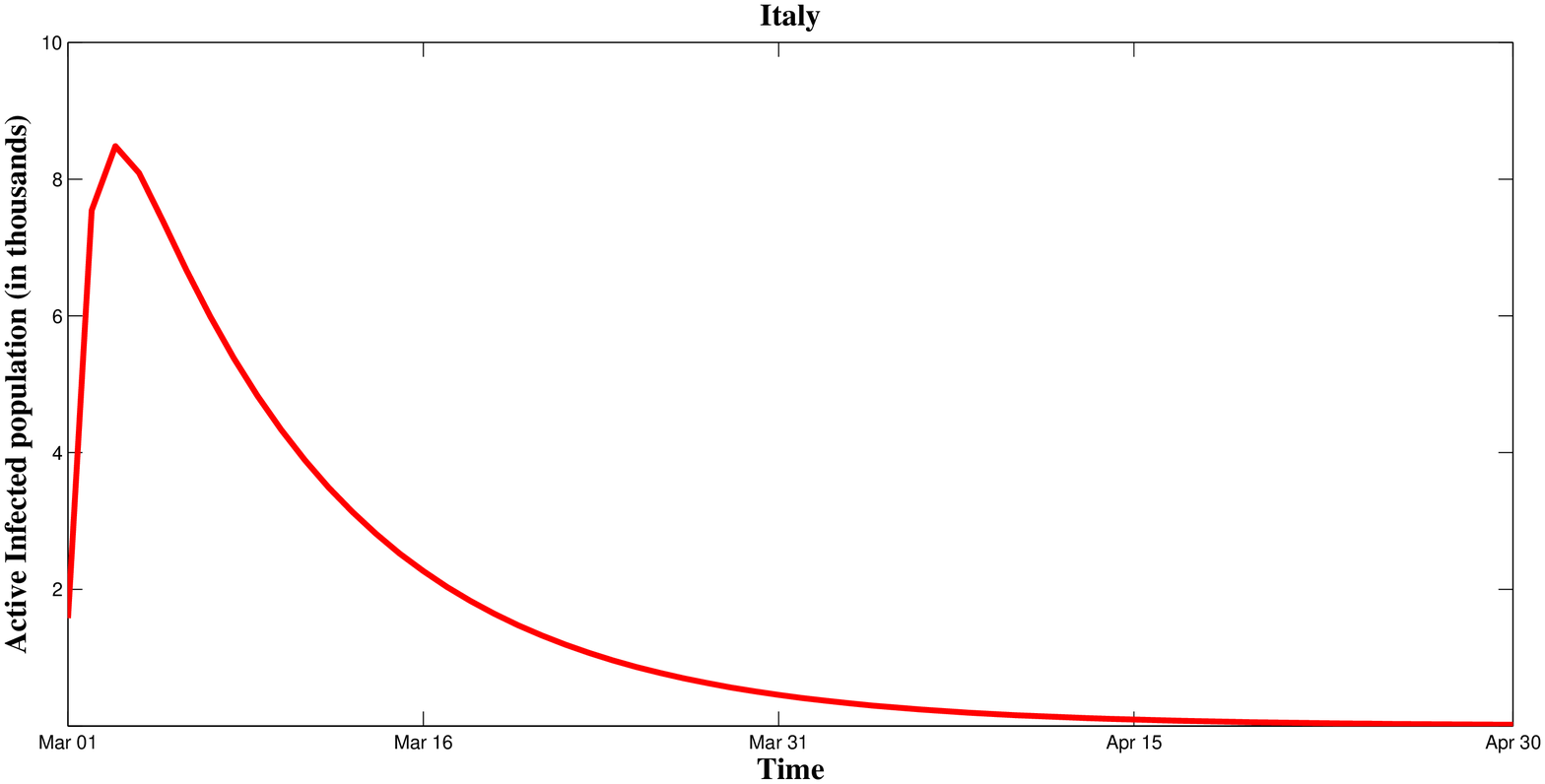}
\caption{$S(0) = 3.5$ Crores}
\label{fig:sub_4b}
\end{subfigure}
\begin{center}

\begin{subfigure}{0.8\textwidth}
\centering
\includegraphics[width=\linewidth]{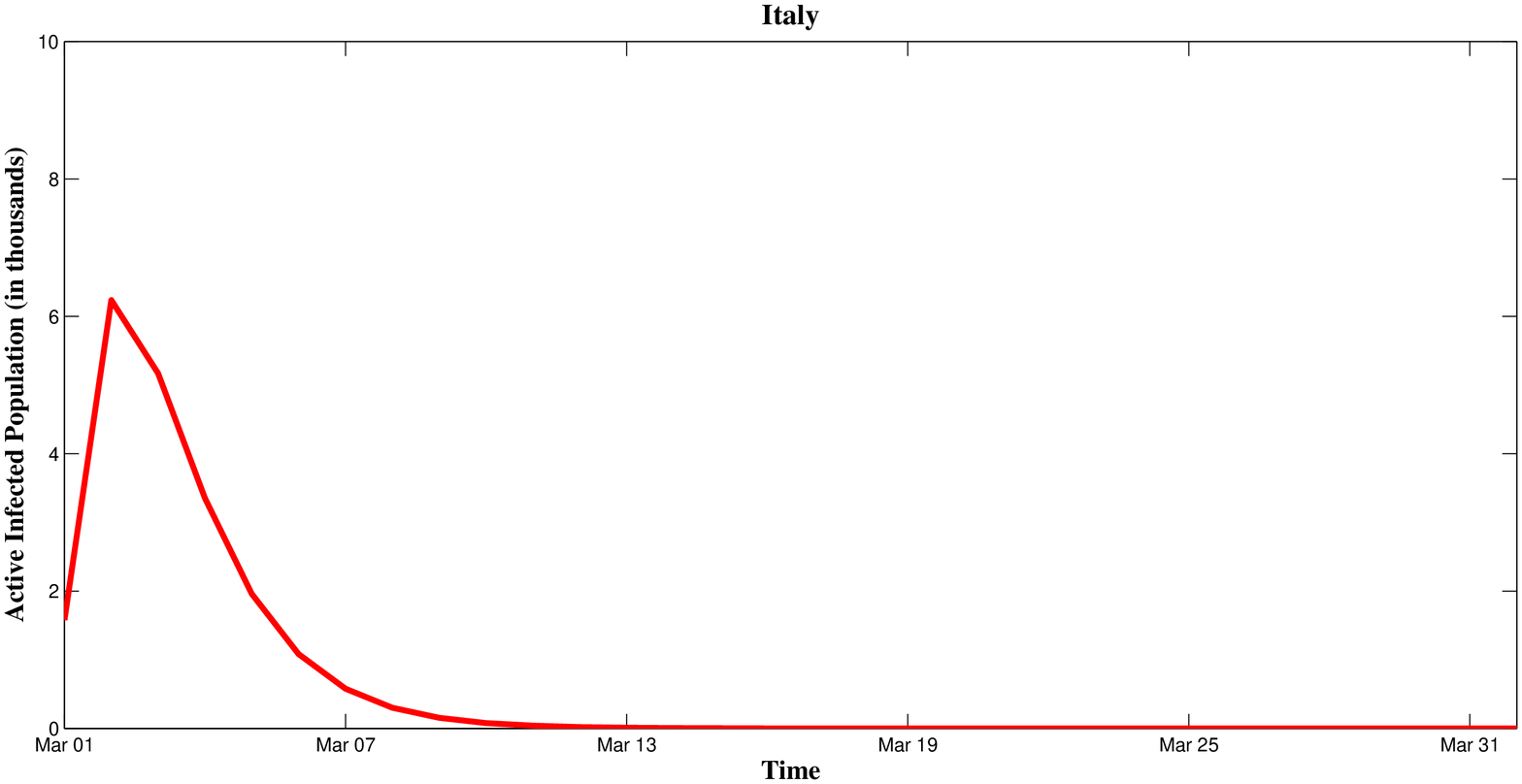}
\caption{$S(0) = 10$ Lakhs}
\label{fig:sub_4c}
\end{subfigure}

\end{center}
\caption{Variation in infected population for different values of initial susceptible population $S(0)$. (a) $S(0)$ = 5 Crores (b) $S(0)$ = 3.5 Crores (c) $S(0)$ = 10 Lakhs. The initial point for simulations has been assumed as March $1^{st}$, 2020.  }
\label{fig_4}
\end{figure}
\\It is clear from Figure \ref{fig_4} that the epidemic could have been controlled at a very early stage if the government had imposed the lock down early in Italy. Figure \ref{fig_4}  shows three different scenarios of the epidemic in Italy with different initial values of susceptible class.\\
Figure (\ref{fig:sub_4a}) shows the current scenario of Italy. However, if lock down would have been imposed prior to $9^{th}$ March 2020, the number of susceptible would have been significantly low. In figure (\ref{fig:sub_4b}), we assume the susceptible population to be 3.5 crores due to the lock down in the country. It can be seen from the plot that, it has not only significantly reduce the number of infections, but also caused the overall death of pandemic by $20^{th}$ April, 2020. Also, figure (\ref{fig:sub_4c}) indicate that the epidemic could have been eliminated by $31^{st}$ March, 2020 if the susceptible population have been reduced to 10 lakhs.
\newpage
\subsubsection{Impact of rapid isolation of infected individuals}
\begin{figure}[h]
\begin{subfigure}{.5\textwidth}
\centering
\includegraphics[width=\linewidth]{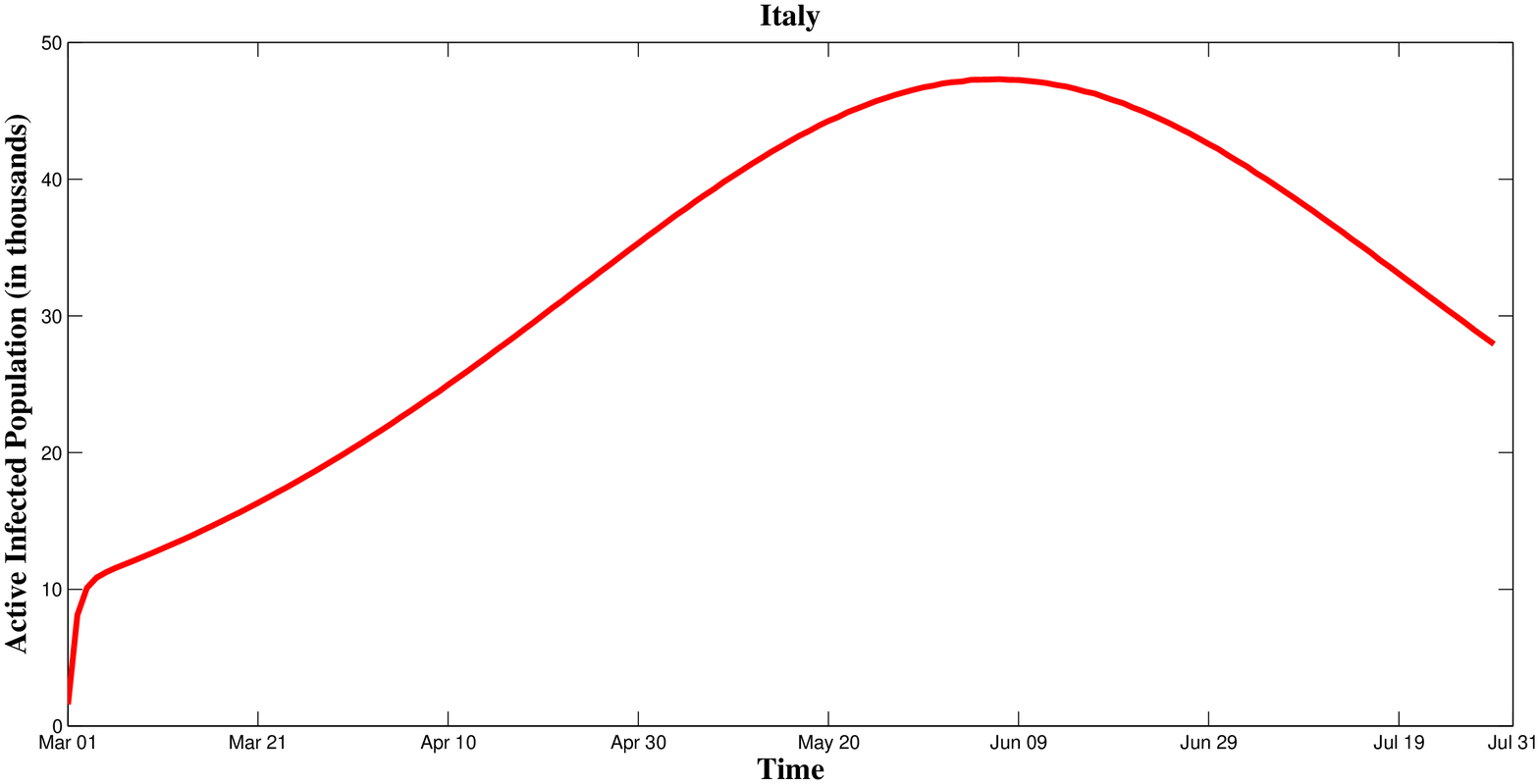}
\caption{$\beta=0.7$}
\label{fig:sub_5a}
\end{subfigure}
\begin{subfigure}{.5\textwidth}
\centering
\includegraphics[width=\linewidth]{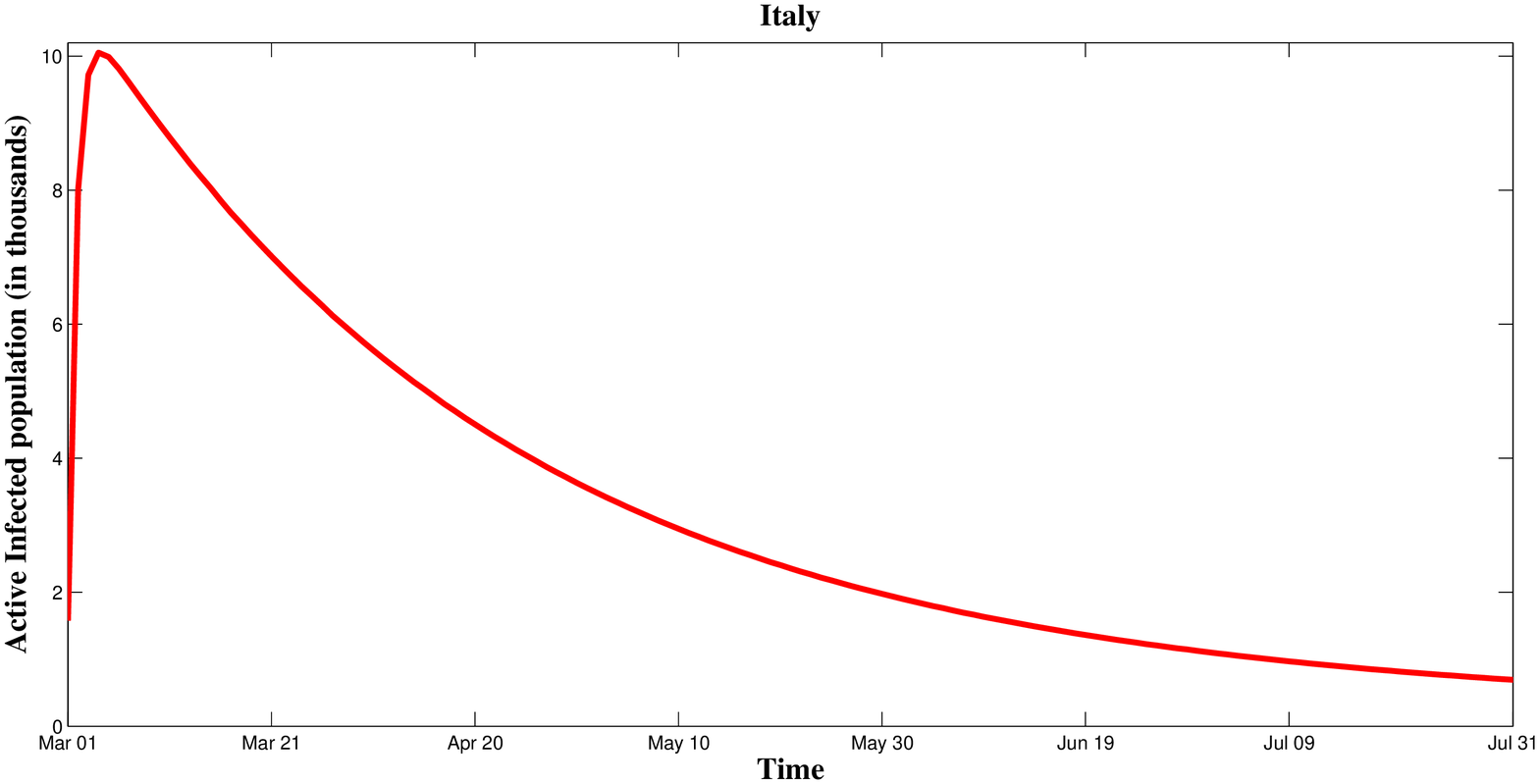}
\caption{$\beta=0.6$}
\label{fig:sub_5b}
\end{subfigure}
\begin{subfigure}{.5\textwidth}
\centering
\includegraphics[width=\linewidth]{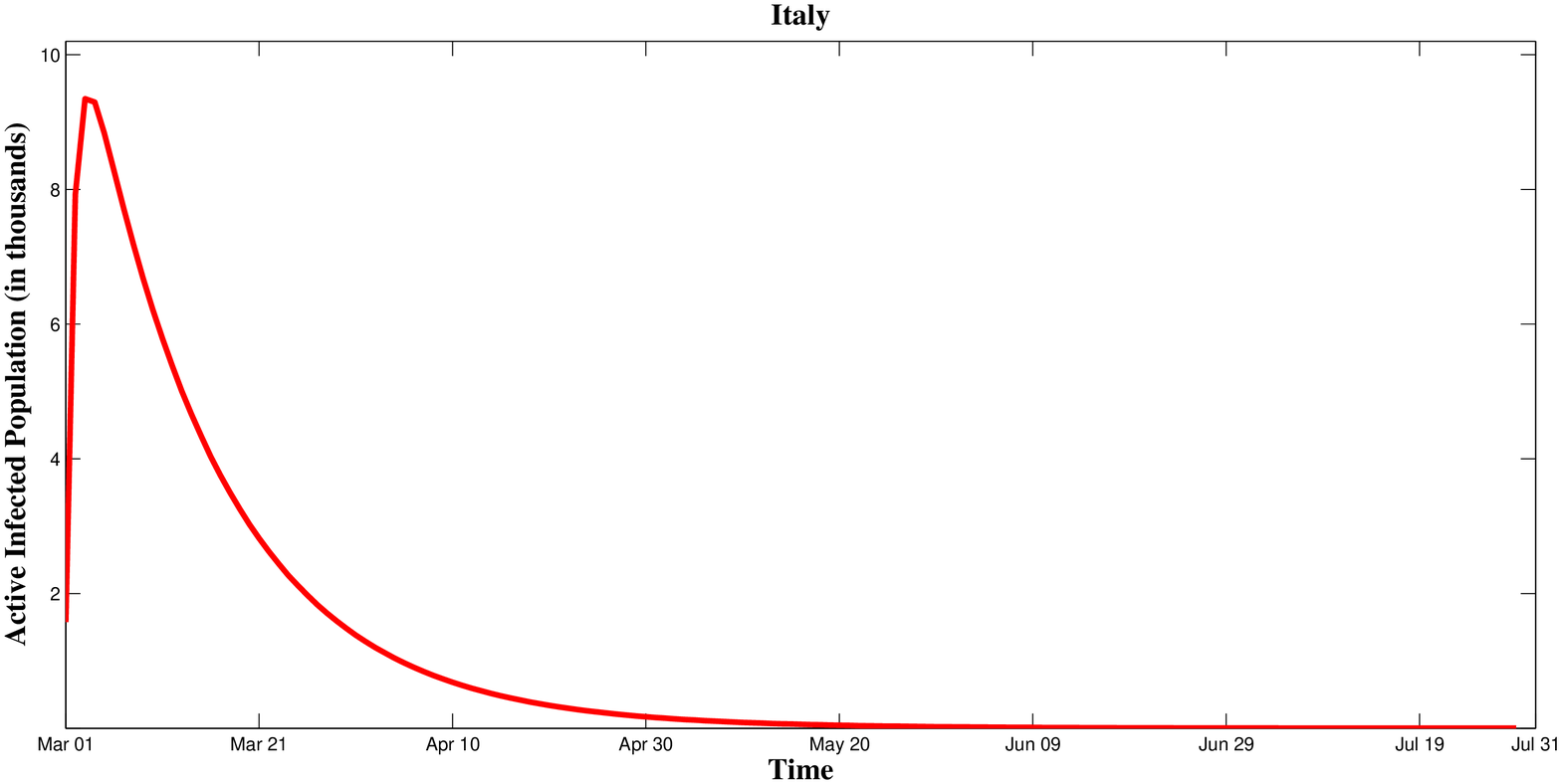}
\caption{$\beta=0.5$}
\label{fig:sub_5c}
\end{subfigure}
\begin{subfigure}{.5\textwidth}
\centering
\includegraphics[width=\linewidth]{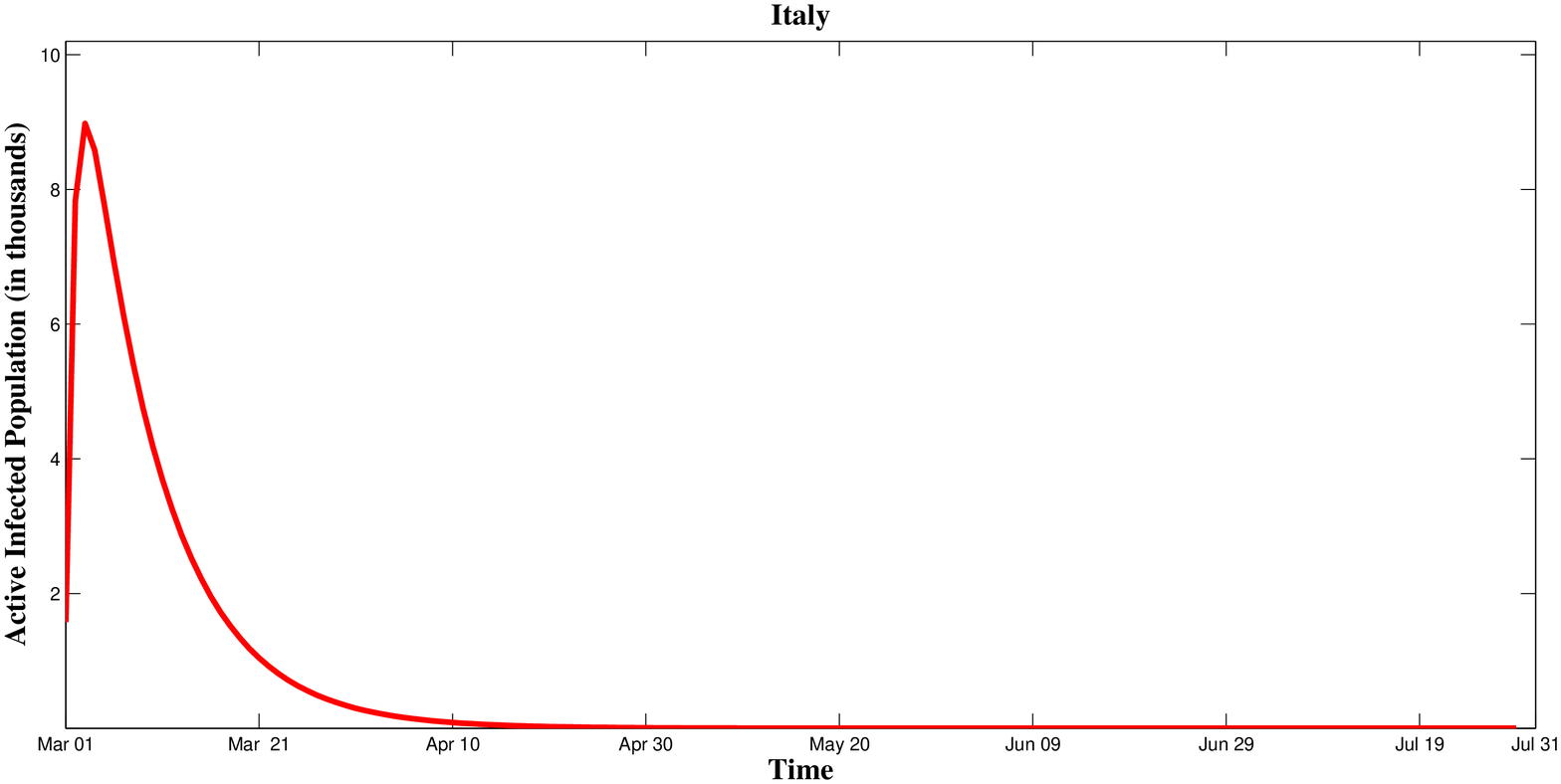}
\caption{$\beta=0.4$}
\label{fig:sub_5d}
\end{subfigure}
\caption{Variation in infected population for different values of $\beta$. (a) $\beta$ = 0.7 (b) $\beta$ = 0.6 (c) $\beta$ = 0.5 (d) $\beta$ =0.4. The initial point for simulation is assumed as March $1$, 2020. }
\label{fig_5}
\end{figure}

COVID-19 is a global pandemic which is spreading all across the globe. Early research shows that the disease transmission rate from an infected individual to a susceptible is very high \cite{liu2020reproductive}. The transmission rate can be reduced by isolating the infected individuals as quickly as possible. In this subsection, with the help of numerical simulations we will show the variations in infected population for different values of $\beta$, disease transmission rate from infected individual to susceptible individuals. \\
Figure \ref{fig_5} shows various scenarios of the epidemic in Italy in case disease transmission rate would have been timely controlled. A rapid isolation of infected population will lead to reduce the disease transmission rate, $\beta$. From figure \ref{fig_5}, we see that as disease transmission rate, $\beta$ is reduced from $0.7$ to $0.4$, it not only decrease the active number of infections from 45000 to 9000, but also the overall lifespan of pandemic reduced from July 31 to April 20, 2020.

\section{Pandemic in different Regions of Italy }
 In this section, we will study and analyze the spread of the pandemic in different regions of Italy. There are few regions which have become hotspots for the pandemic in the country. These regions include Lombardia, Emilia Romagna, Piemonte and Veneto. In order to study the pandemic more precisely, we have assumed that the transmission rate in these regions varies from the generalized transmission rate of Italy. This can be explained, as there is a visible population gap in these regions. Also, we have assumed that by first week of March, whole population of these regions is susceptible. Hence, we have assumed that there are no new individuals available for recruitments, hence $A$ equals zero. For this purpose we have used the parameters values available in Table \ref{table2}. The disease  transmission rate from exposed individuals is also reduced to 0.4. All the remaining parameter values are considered to be same as given in Table \ref{Table: 1}.  \\

\begin{table}[h]
\begin{tabular}{|p{0.8 cm}|p{.8 in}|p{.7 in}|p{.8 in}|p{.7 in}|p{.7 in}|}\hline
Sr. No. & Lambardia & Emilia Romegna & Piemonte & Veneto & Source\\ \hline
$N$ & 10060574 & 4459477 & 4356406 & 4905854 & \cite{statista} \\ \hline
$\beta$& 0.7 & 0.68 & 0.68 & 0.69 & Assumed\\ \hline
\end{tabular}
\caption{Parameter values for regions of Italy}
\label{table2}
\end{table}

\begin{figure}[h]
\begin{subfigure}{.5\textwidth}
\centering
\includegraphics[width=\linewidth]{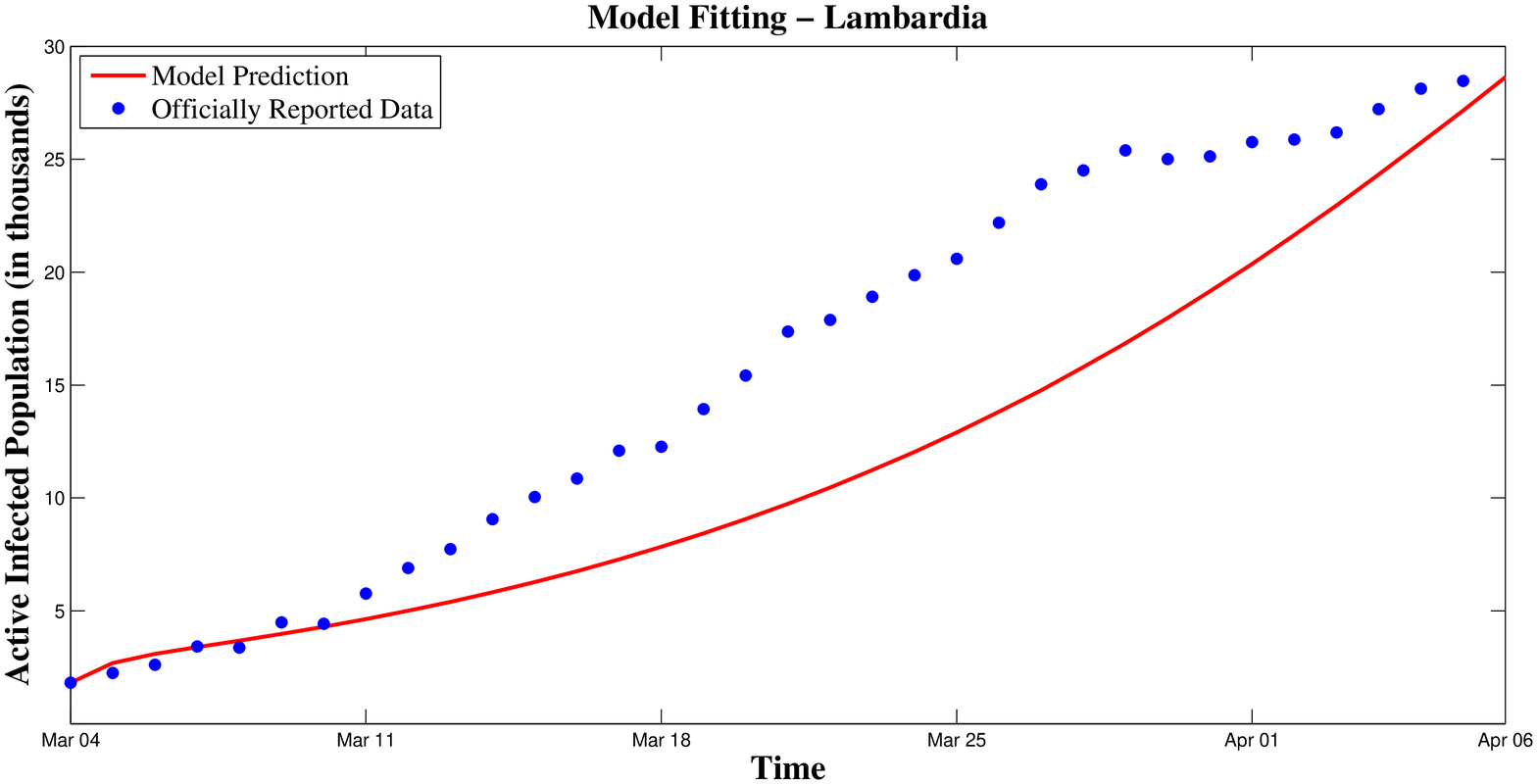}
\caption{}
\label{fig:sub_6a}
\end{subfigure}
\begin{subfigure}{.5\textwidth}
\centering
\includegraphics[width=\linewidth]{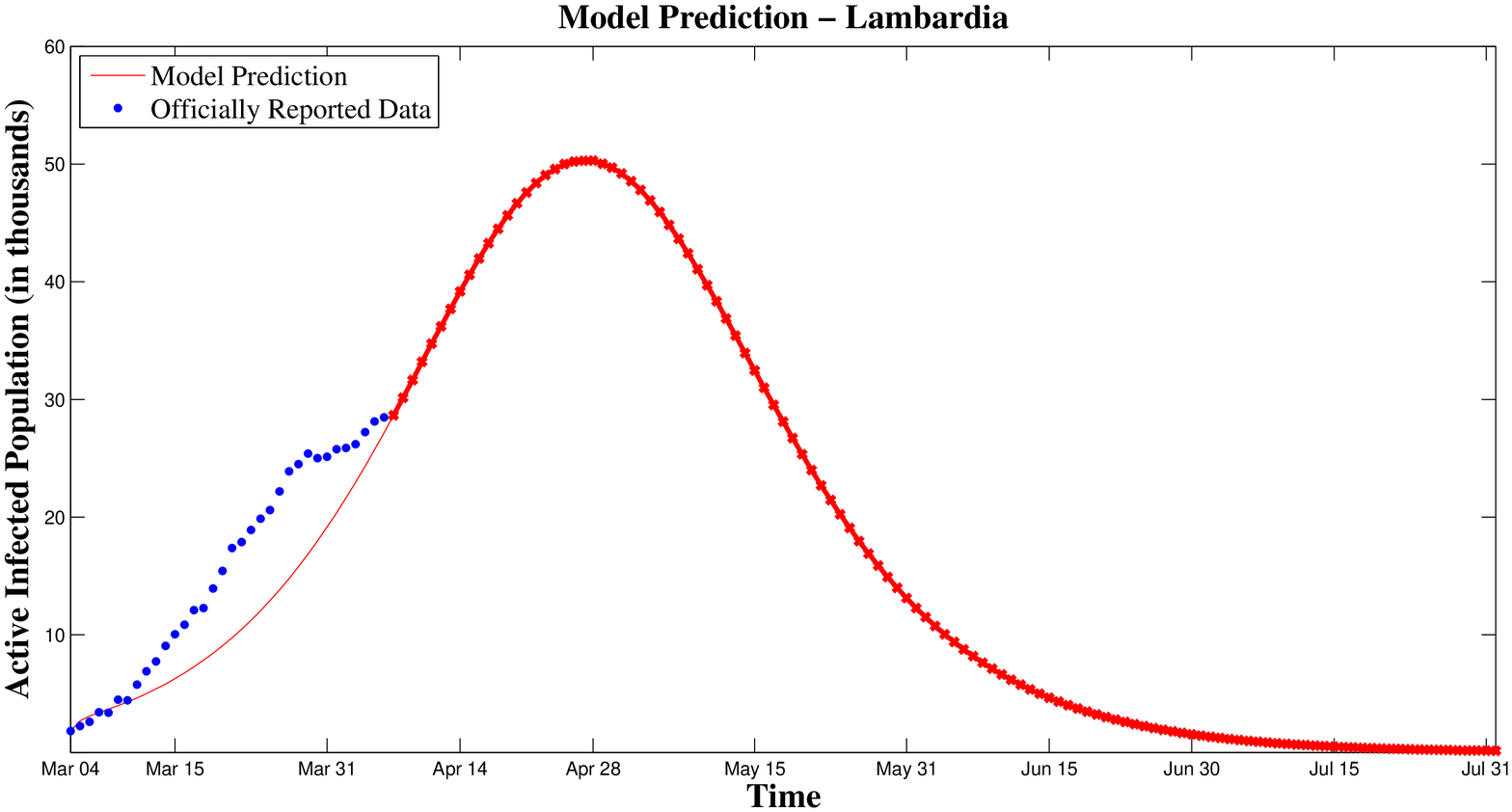}
\caption{}
\label{fig:sub_6b}
\end{subfigure}
\caption{Case study of Lambardia. (a) Model fitting from $4^{th}$ March 2020 to $6^{th}$ April 2020. (b) Future of COVID-19 using model simulation. }
\label{fig_6}
\end{figure}

\begin{figure}[h]
\begin{subfigure}{.5\textwidth}
\centering
\includegraphics[width=\linewidth]{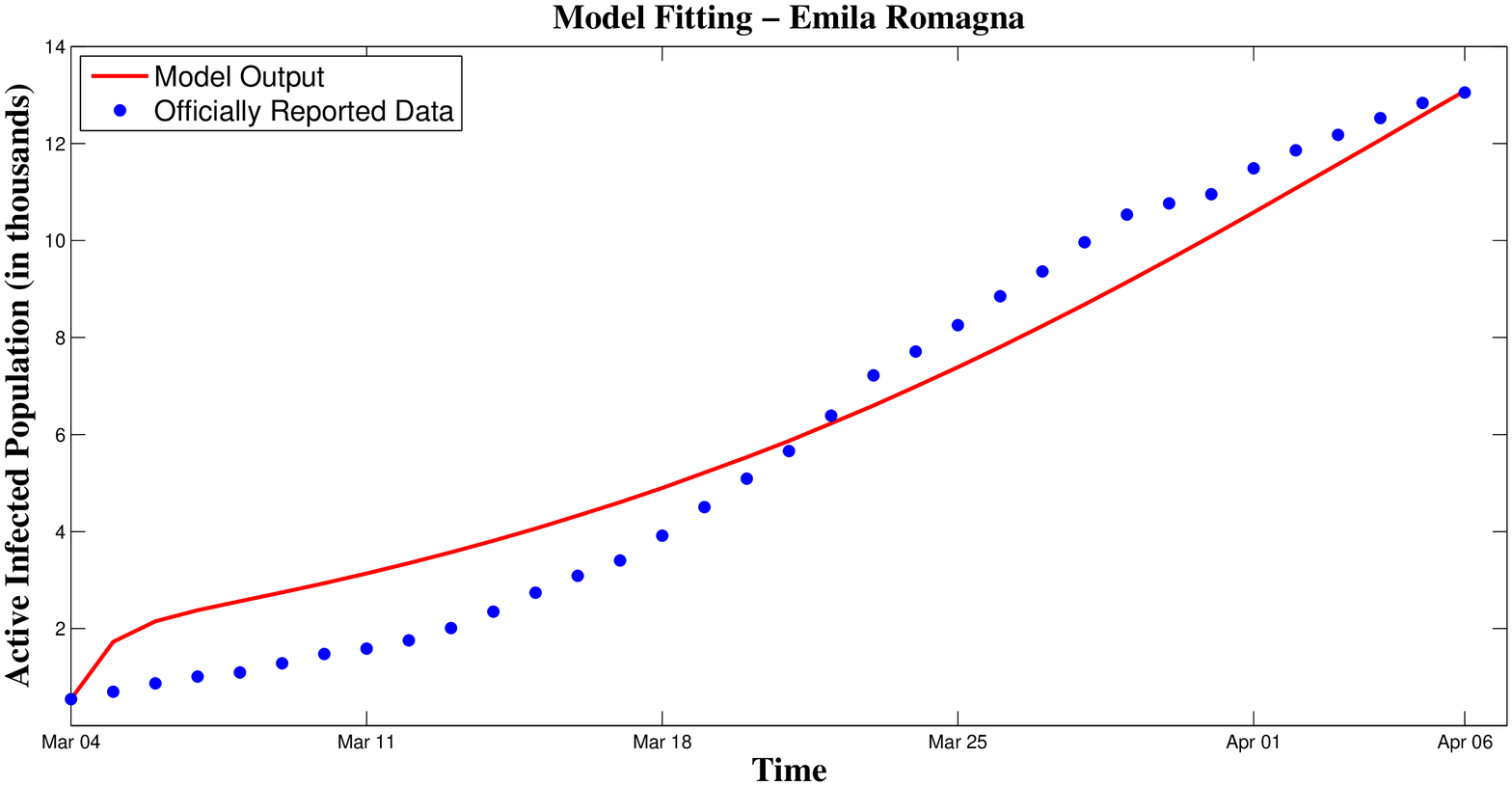}
\caption{}
\label{fig:sub_7a}
\end{subfigure}
\begin{subfigure}{.5\textwidth}
\centering
\includegraphics[width=\linewidth]{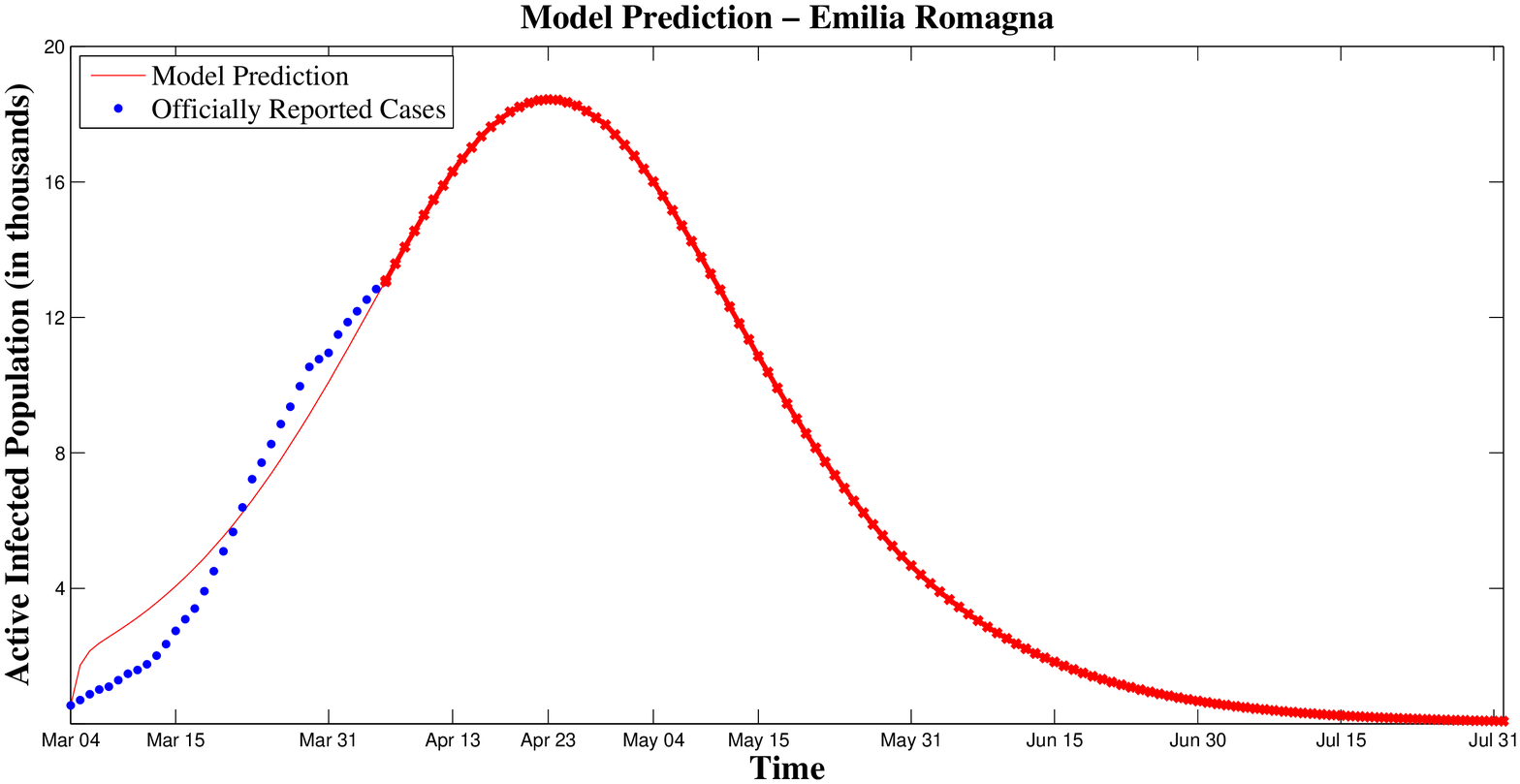}
\caption{}
\label{fig:sub_7b}
\end{subfigure}
\caption{Case study of Emilia Romagna. (a) Model fitting from $4^{th}$ March 2020 to $6^{th}$ April 2020. (b) Future of COVID-19 using model simulation.  }
\label{fig_7}
\end{figure}

\begin{figure}[h]
\begin{subfigure}{.5\textwidth}
\centering
\includegraphics[width=\linewidth]{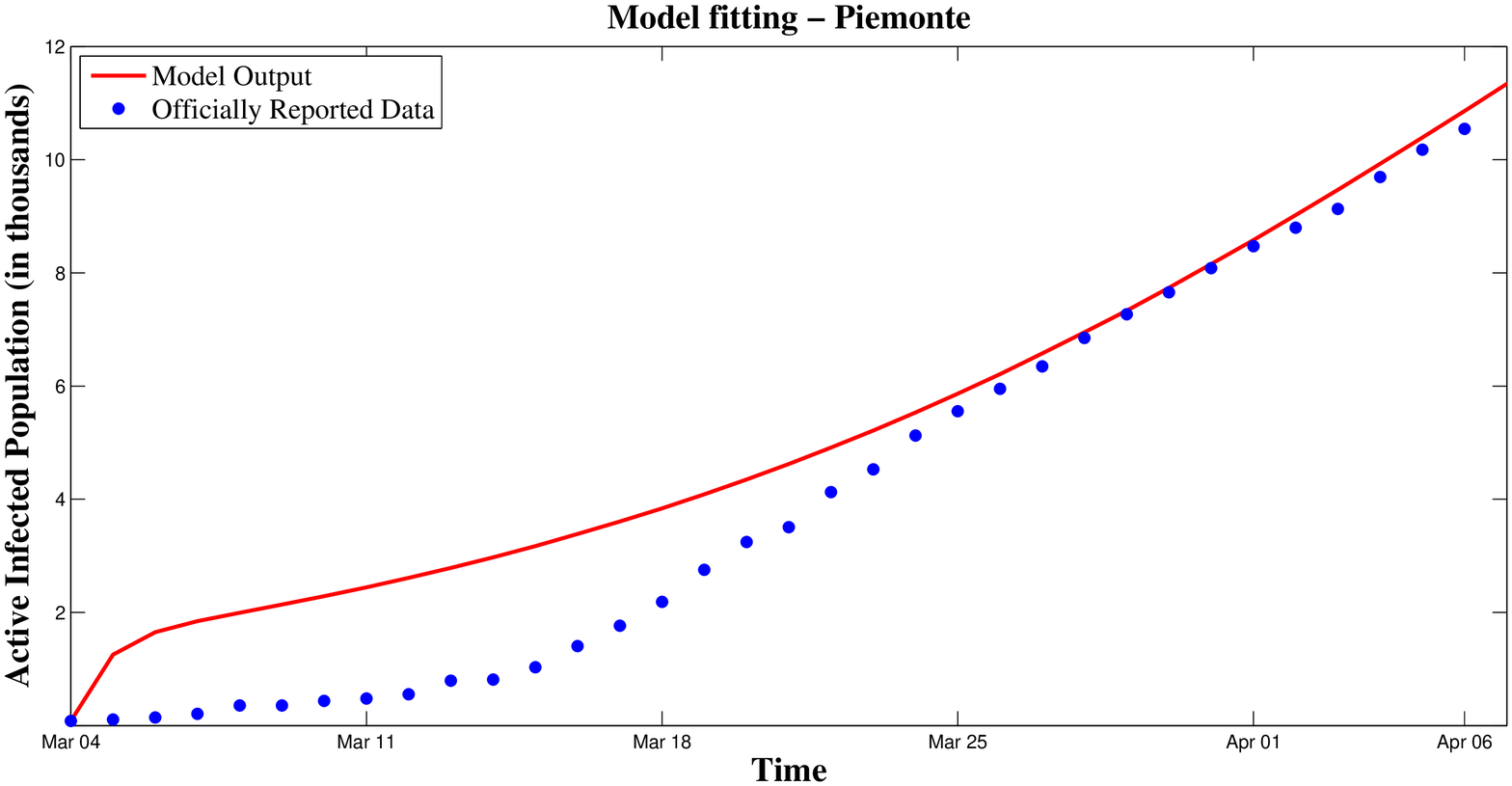}
\caption{}
\label{fig:sub_8a}
\end{subfigure}
\begin{subfigure}{.5\textwidth}
\centering
\includegraphics[width=\linewidth]{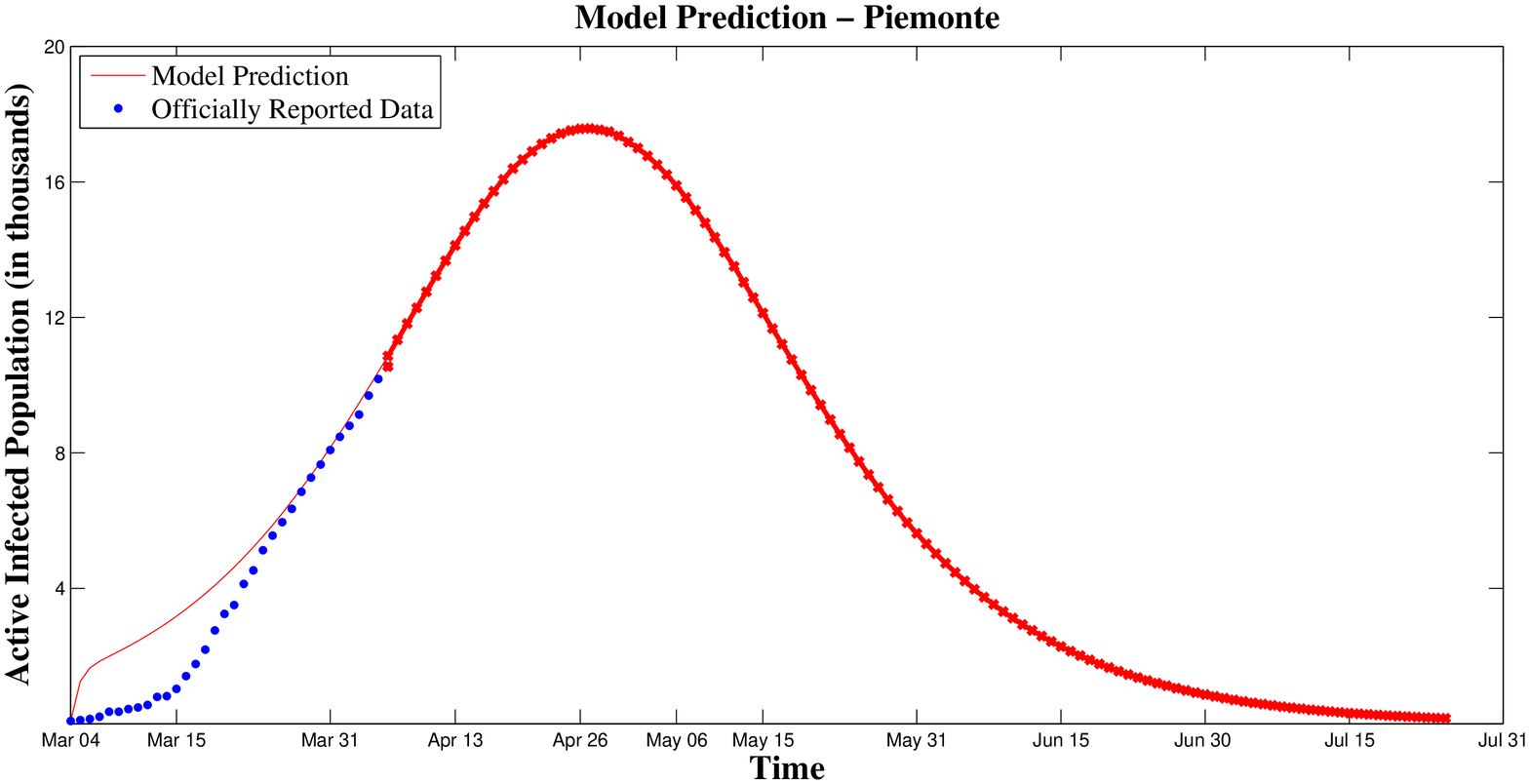}
\caption{}
\label{fig:sub_8b}
\end{subfigure}
\caption{Case study of Piemonte. (a) Model fitting from $4^{th}$ March 2020 to $6^{th}$ April 2020. (b) Future of COVID-19 using model simulation.  }
\label{fig_8}
\end{figure}

\begin{figure}[h]
\begin{subfigure}{.5\textwidth}
\centering
\includegraphics[width=\linewidth]{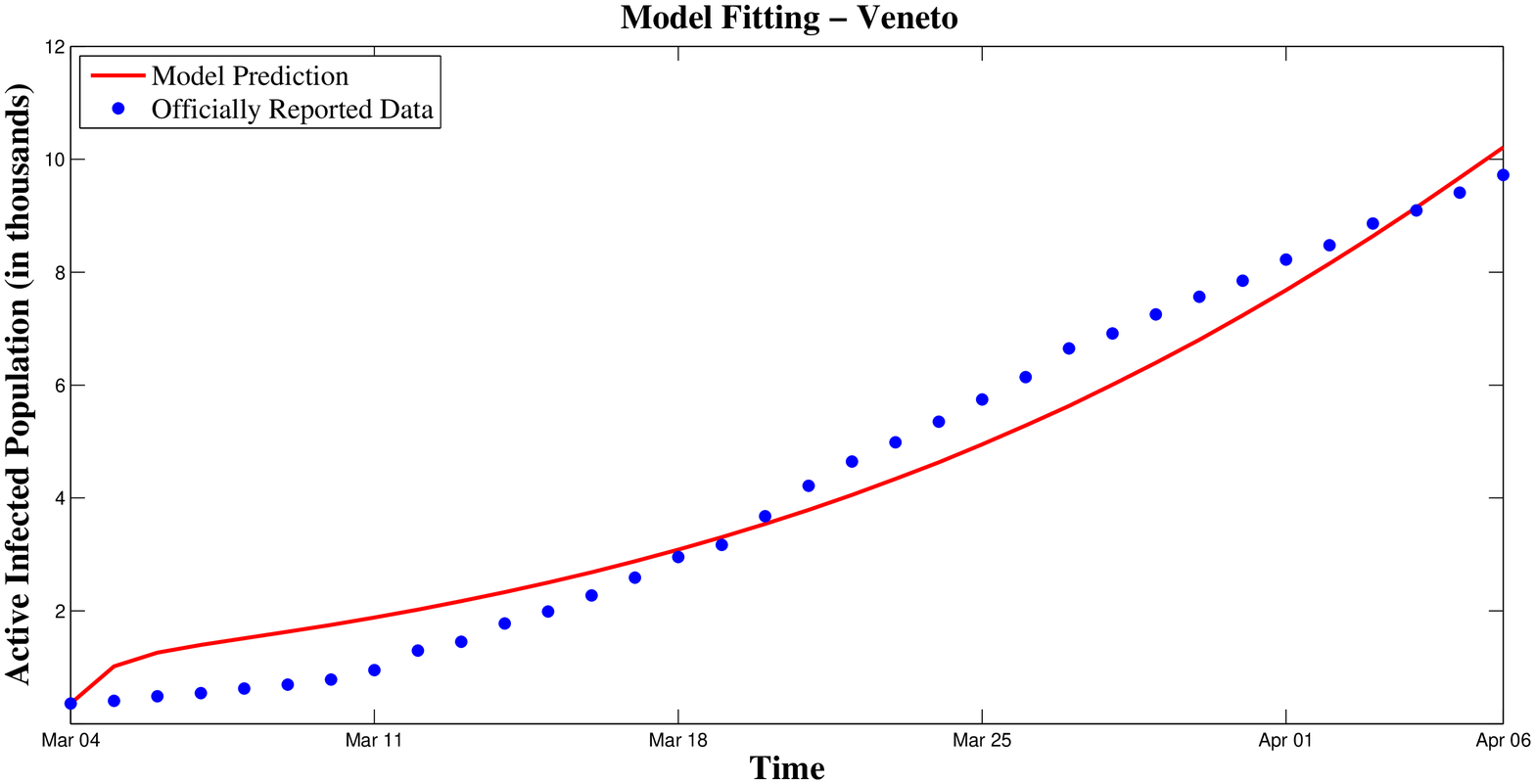}
\caption{}
\label{fig:sub_9a}
\end{subfigure}
\begin{subfigure}{.5\textwidth}
\centering
\includegraphics[width=\linewidth]{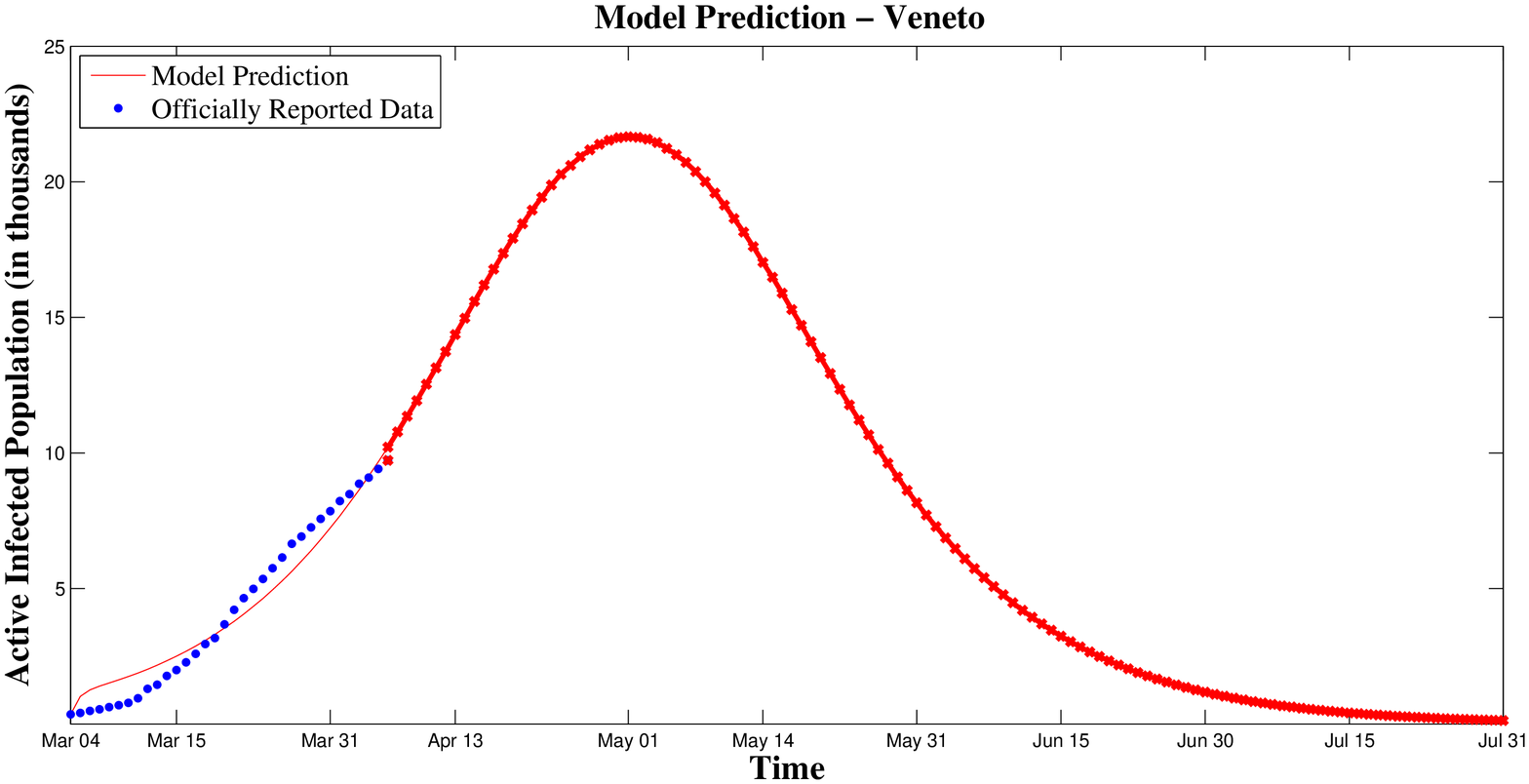}
\caption{}
\label{fig:sub_9b}
\end{subfigure}
\caption{Case study of Veneto. (a) Model fitting from $4^{th}$ March 2020 to $6^{th}$ April 2020. (b) Future of COVID-19 using model simulation.  }
\label{fig_9}
\end{figure}

In figure \ref{fig_6}, we study the case of Lambardia region of Italy. It is interesting to note that cases of infectives were highest in this region of the country. In sub-figure \ref{fig:sub_6a}, the proposed model is used for fitting the available data. In sub-figure \ref{fig:sub_6b}, we used our proposed model to predict the future of pandemic in Lambardia. We see from the study that by mid July, the pandemic will die out in this region. Figure \ref{fig_7} shows the case study of Emilia Romagna region of Italy. In figure (\ref{fig:sub_7a}), we can see the accuracy of our proposed model. Figure (\ref{fig:sub_7b}) consist of the upcoming predicted scenario of the epidemic in this region. According to our proposed model the pandemic will last till mid July, 2020 in this particular region.\\
Figures \ref{fig_8} and \ref{fig_9} illustrate the behaviour of the pandemic in Piemonte and Veneto regions respectively. Sub-figures (\ref{fig:sub_9a}) and (\ref{fig:sub_10a}) exhibit the authenticity of our proposed model with the officially reported COVID-19 cases in these regions. The possible extinction of pandemic in these regions is shown in sub-figures (\ref{fig:sub_8a}) and (\ref{fig:sub_9a}). According to our proposed model, the pandemic in both of these regions will last till the last week of July, 2020. Also, it can be observed from sub-figure (\ref{fig:sub_8b}) that the active infected cases in Piemonte will not surpass 18000. Whereas, the active infected cases in Veneto may hit the 21000 mark ( see sub-fig \ref{fig:sub_9b}).

 Figures \ref{fig_6}, \ref{fig_7}, \ref{fig_8} and \ref{fig_9} shows the accuracy of our model with the officially reported data of these regions. Also, from our simulations, one can conclude that the epidemic is yet to reach its peak.

\newpage
\section{Basic Reproduction Number}
\begin{figure}[h]
\begin{subfigure}{.5\textwidth}
\centering
\includegraphics[width=\linewidth]{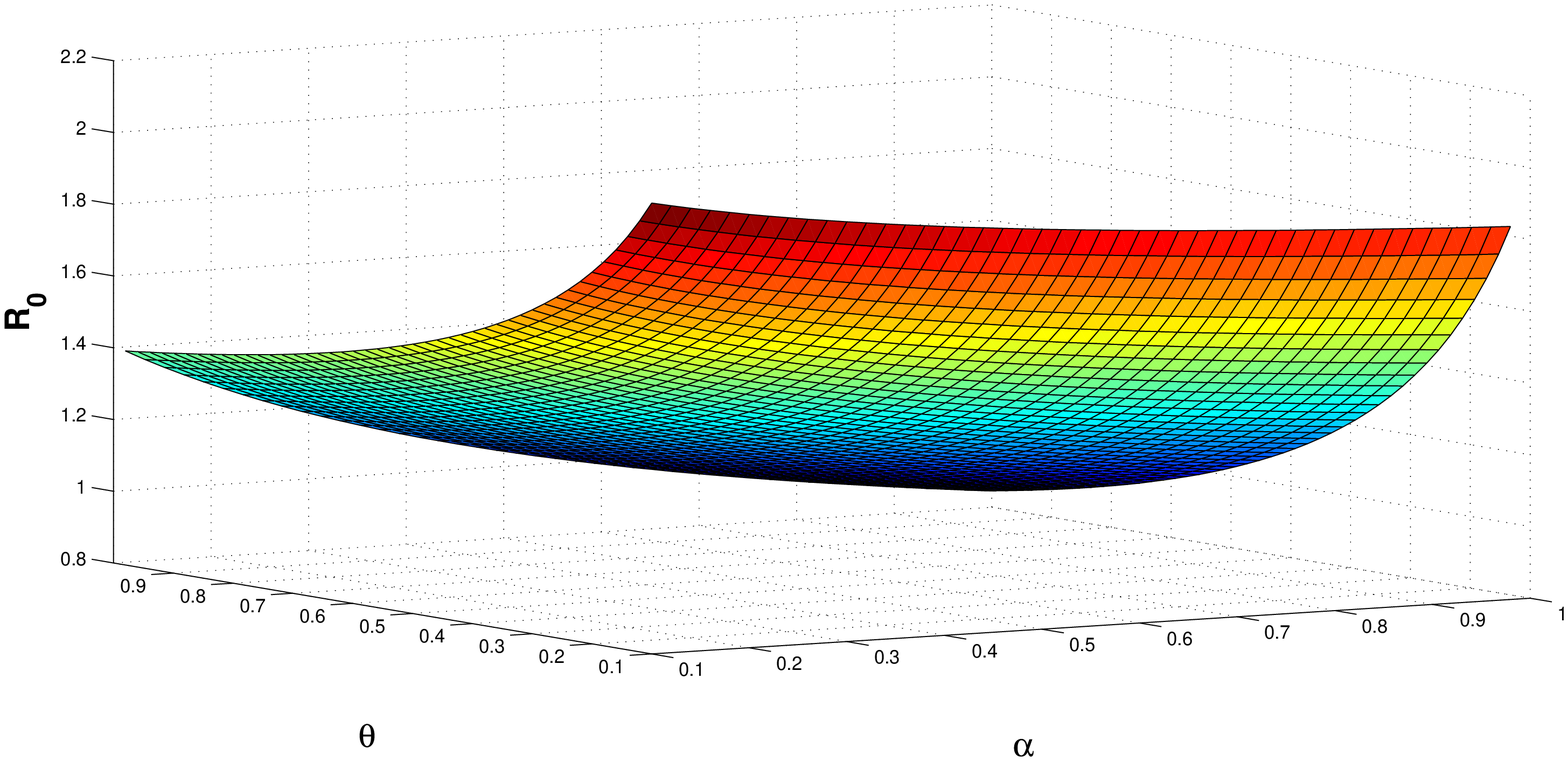}
\caption{}
\label{fig:sub_10a}
\end{subfigure}
\begin{subfigure}{.5\textwidth}
\centering
\includegraphics[width=\linewidth]{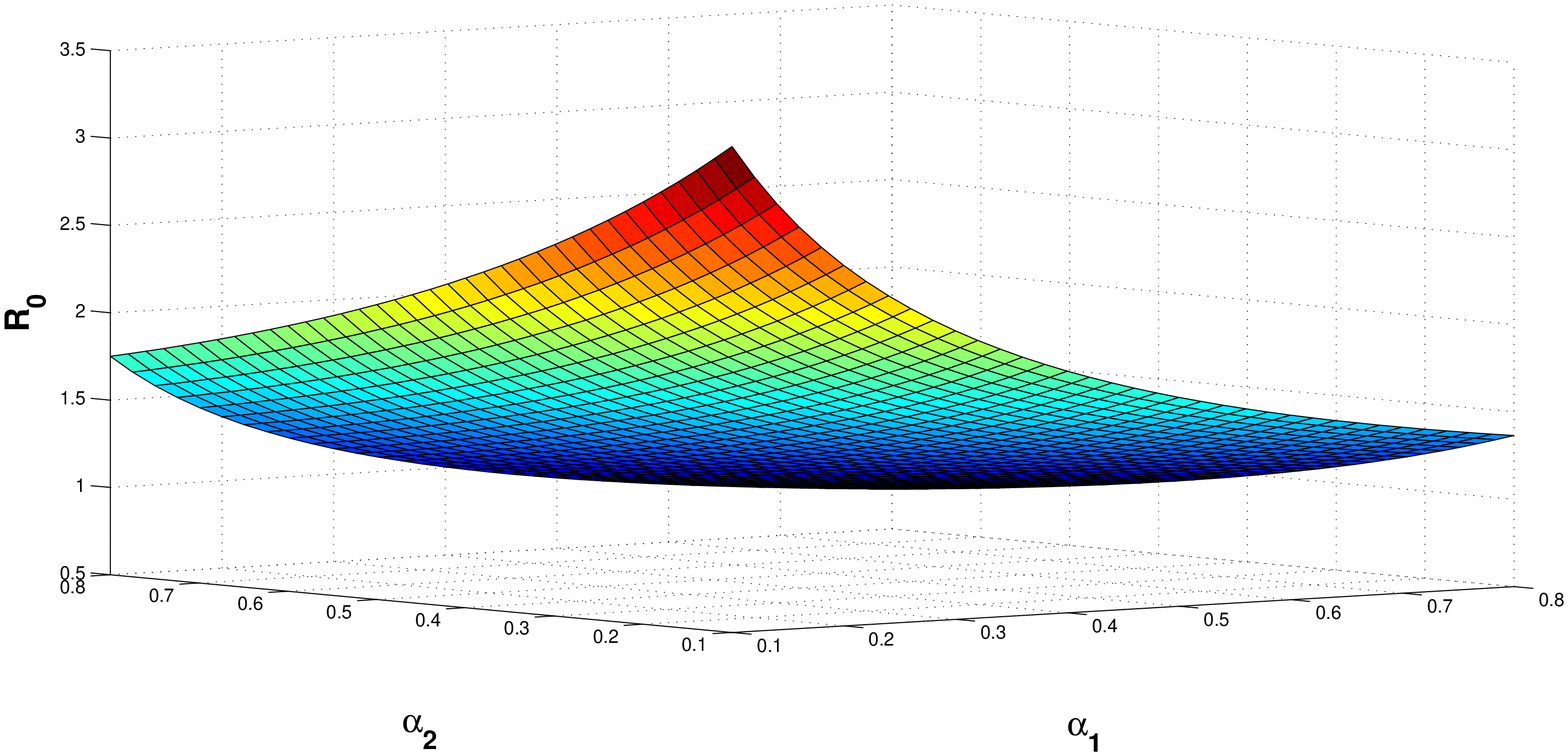}
\caption{}
\label{fig:sub_10b}
\end{subfigure}

\begin{subfigure}{.5\textwidth}
\centering
\includegraphics[width=\linewidth]{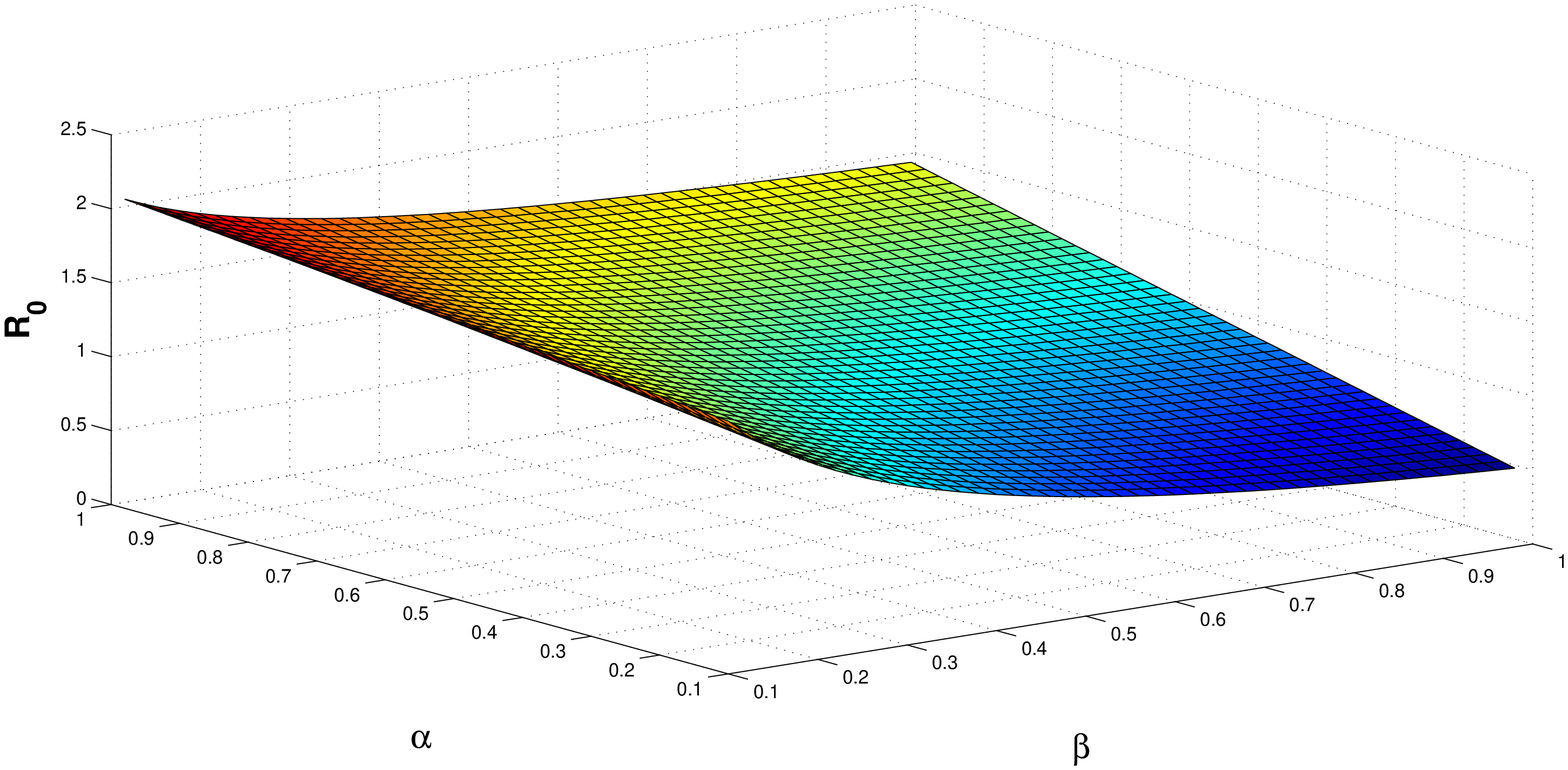}
\caption{}
\label{fig:sub_10c}
\end{subfigure}
\begin{subfigure}{.5\textwidth}
\centering
\includegraphics[width=\linewidth]{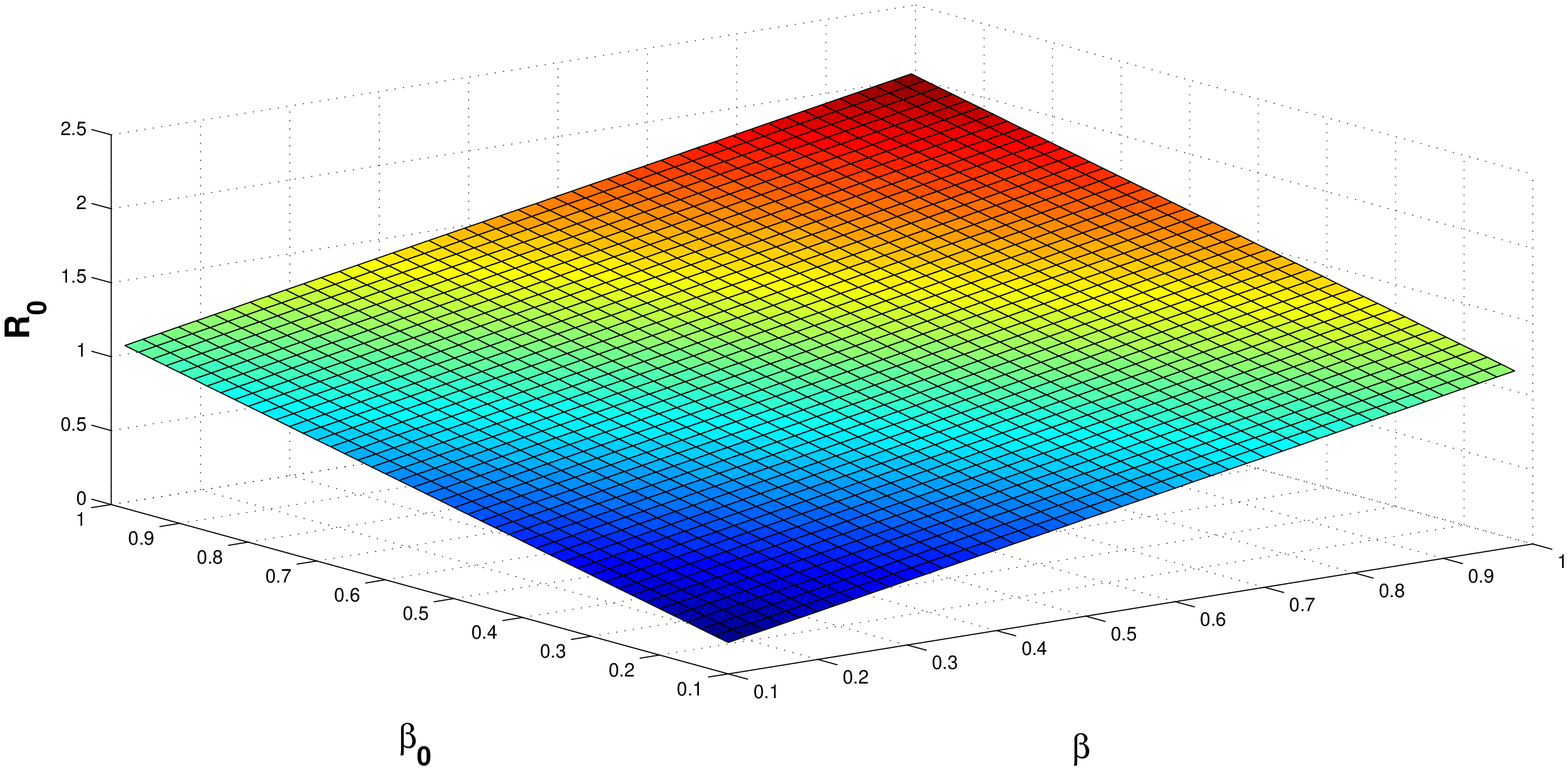}
\caption{}
\label{fig:sub_10d}
\end{subfigure}
\caption{Variation in the basic reproduction number $R_0$  for different values of sensitive parameters. (a) Effect of $\alpha$ and $\theta$ on $R_0$. (b) Effect of $\alpha_1$ and $\alpha_2$ on $R_0$. (c) Effect of $\beta$ and $\alpha$ on $R_0$. (d) Effect of $\beta$ and $\beta_0$ on $R_0$.}
\label{fig_10}
\end{figure}

The basic reproduction number is one of the key parameter to examine the long term behaviour of an epidemic. It can be defined as the number of secondary cases produced by a single infected individual in its entire life span as infectious agent. We have used next-generation matrix technique explained in \cite{diekmann2010construction}, to derive the expression of basic reproduction number $R_0$. The $R_0$ takes the following expression

\[R_0=\frac{\beta_0\left(\alpha_2+\theta+\mu\right)+\beta \alpha}{\left(\alpha+\alpha_1+\mu\right)\left(\alpha_2+\theta+\mu\right)}\]
We will analyse the variation in $R_0$ for different values of the parameters involved in the model system. Figure \ref{fig_10} illustrates the simultaneous variation in the basic reproduction number for different values of corresponding parameters. The parameter values used are given in Table \ref{Table: 1}.

\newpage
\section{Conclusion}
A SEIR type compartmental model is proposed to study the current scenario of COVID-19 in Italy. Our proposed model accurately fits the officially available data of the pandemic in Italy. Also, it is concluded from the study  that the pandemic may still grow in Italy till $30^{th}$April, 2020. The pandemic may hit the 2.5 lacks active cases mark (see fig. \ref{fig_3}) in Italy. Also, we have discussed how the lock down that was imposed on $9^{th}$ March, 2020 was a good but a delayed decision of the government of Italy. Through simulations, we have shown that a rapid isolation of the infective individuals and early lock down in the country are two of the most efficient procedures to terminate the spread of COVID-19. As of now, the vaccination of COVID-19 have not been discovered. Hence, this research can also be beneficial for the countries which are in the initial stage of the pandemic, as our research describes two of the most effective procedures to counter the spread of the pandemic and its long term impact on the spread of disease. \\
The expression of the basic reproduction number $R_0$ is also derived. As our proposed model involves various parameters, we have shown the sensitivity of these parameters via numerical simulations. It is clear from the simulations (see Fig. \ref{fig:sub_10d}) that the transmission rates, $\beta $ and $\beta_0$  are the most sensitive parameters. The reproduction number can be minimized if we can reduce these two parameters.\\
This research can be extended in various ways. One can refine the model by introducing new compartments in order to examine the epidemic more precisely. There are certain assumptions which we have made while constructing this model because of the limited data and short onset time. As more data will be available in the future, this model can be trained with more real data to increase its efficiency.   
\section*{References}

\begin{thebibliography}{10}
\expandafter\ifx\csname url\endcsname\relax
  \def\url#1{\texttt{#1}}\fi
\expandafter\ifx\csname urlprefix\endcsname\relax\def\urlprefix{URL }\fi
\expandafter\ifx\csname href\endcsname\relax
  \def\href#1#2{#2} \def\path#1{#1}\fi

\bibitem{lai2020severe}
C.-C. Lai, T.-P. Shih, W.-C. Ko, H.-J. Tang, P.-R. Hsueh, Severe acute
  respiratory syndrome coronavirus 2 (sars-cov-2) and corona virus disease-2019
  (covid-19): the epidemic and the challenges, International journal of
  antimicrobial agents (2020) 105924.

\bibitem{lupia20202019}
T.~Lupia, S.~Scabini, S.~M. Pinna, G.~Di~Perri, F.~G. De~Rosa, S.~Corcione,
  2019-novel coronavirus outbreak: A new challenge, Journal of Global
  Antimicrobial Resistance (2020).

\bibitem{who:2020}
{\url{https://www.who.int/emergencies/diseases/novel-coronavirus-2019}}.

\bibitem{who1:2020}
{\url{https://www.who.int/emergencies/diseases/novel-coronavirus-2019/events-as-they-happen}}.

\bibitem{li2020early}
Q.~Li, X.~Guan, P.~Wu, X.~Wang, L.~Zhou, Y.~Tong, R.~Ren, K.~S. Leung, E.~H.
  Lau, J.~Y. Wong, et~al., Early transmission dynamics in wuhan, china, of
  novel coronavirus--infected pneumonia, New England Journal of Medicine
  (2020).

\bibitem{wang2020clinical}
D.~Wang, B.~Hu, C.~Hu, F.~Zhu, X.~Liu, J.~Zhang, B.~Wang, H.~Xiang, Z.~Cheng,
  Y.~Xiong, et~al., Clinical characteristics of 138 hospitalized patients with
  2019 novel coronavirus--infected pneumonia in wuhan, china, Jama (2020).

\bibitem{carlos2020novel}
W.~G. Carlos, C.~S. Dela~Cruz, B.~Cao, S.~Pasnick, S.~Jamil, Novel wuhan
  (2019-ncov) coronavirus, American journal of respiratory and critical care
  medicine 201~(4) (2020) P7--P8.

\bibitem{biscayart2020next}
C.~Biscayart, P.~Angeleri, S.~Lloveras, T.~d. S.~S. Chaves, P.~Schlagenhauf,
  A.~J. Rodr{\'\i}guez-Morales, et~al., The next big threat to global health?
  2019 novel coronavirus (2019-ncov): What advice can we give to
  travellers?--interim recommendations january 2020, from the latin-american
  society for travel medicine (slamvi) (2020).

\bibitem{grasselli2020critical}
G.~Grasselli, A.~Pesenti, M.~Cecconi, Critical care utilization for the
  covid-19 outbreak in lombardy, italy: early experience and forecast during an
  emergency response, JAMA (2020).

\bibitem{remuzzi2020covid}
A.~Remuzzi, G.~Remuzzi, Covid-19 and italy: what next?, The Lancet (2020).

\bibitem{lazzerini2020covid}
M.~Lazzerini, G.~Putoto, Covid-19 in italy: momentous decisions and many
  uncertainties, The Lancet Global Health (2020).

\bibitem{vattay2020predicting}
G.~Vattay, Predicting the ultimate outcome of the covid-19 outbreak in italy,
  arXiv preprint arXiv:2003.07912 (2020).

\bibitem{worldometer}
{https://www.worldometers.info/coronavirus/country/italy/}.

\bibitem{Knoema}
https://knoema.com/legal/termsofuse (April 2020).
\newblock \href{https://knoema.com/legal/termsofuse}{[link]}.
\newline\urlprefix\url{https://knoema.com/legal/termsofuse}

\bibitem{situation}
https://www.who.int/emergencies/diseases/novel-coronavirus-2019/situation-reports.

\bibitem{liu2020reproductive}
Y.~Liu, A.~A. Gayle, A.~Wilder-Smith, J.~Rockl{\"o}v, The reproductive number
  of covid-19 is higher compared to sars coronavirus, Journal of travel
  medicine (2020).

\bibitem{statista}
{https://www.statista.com/statistics/617497/resident-population-italy-by-region/}
  {2020},

\bibitem{diekmann2010construction}
O.~Diekmann, J.~Heesterbeek, M.~G. Roberts, The construction of next-generation
  matrices for compartmental epidemic models, Journal of the Royal Society
  Interface 7~(47) (2010) 873--885.

\end{thebibliography}

\end{document}